\documentclass[12pt]{iopart}
\usepackage{iopams}  
\usepackage{graphicx}
\begin{document}

\title{Equation of State and Transport Coefficients of Relativistic Nuclear
Matter}

\author{Azwinndini Muronga$^{1,2}$}

\address{$^1$ Centre for Theoretical Physics and Astrophysics, Department of
Physics, University of Cape Town, Rondebosch 7701, Cape Town, South Africa.}

\address{$^2$ UCT-CERN Research Centre, Department of Physics, 
University of Cape Town, Rondebosch 7701, Cape Town, South Africa. }

\begin{abstract}

In order to evaluate qualitatively the space-time evolution of hot and dense
nuclear matter the underlying equation of state and transport coefficients must
be known. In this study a specific equation of state is studied: the pion gas.
The classical or standard transport coefficients, namely the bulk viscosity,
shear viscosity and thermal conductivity are divided by the relaxation times for
the corresponding dissipative fluxes and then studied as a function of mass to
temperature ratio.
\end{abstract}

%Uncomment for PACS numbers title message
%\pacs{00.00, 20.00, 42.10}

% Uncomment for Submitted to journal title message
%\submitto{\JPA}

% Comment out if separate title page not required
%\maketitle

\section{Introduction}
In the heavy-ion experiments such as those at the Relativistic Heavy Ion
Collider (RHIC) and the Large Hadron Collider (LHC) a hot and dense matter is
created. The dynamical evolution of such matter proceeds through stages
consisting of both sub-hadronic (quarks and gluons) and hadronic (mesons and
baryons) degrees of freedom. A knowledge of the equation of state and transport
coefficients or relaxation times of various dissipative processes is essential 
for a complete description. For the matter produced at RHIC and LHC one would
like to extract the equation of state and the transport coefficients (and/or the
associated time/length scales) for a given model of the interacting matter.
Then one can study the sensitivity of the space-time evolution of the system and
the calculated distributions of the hadrons to the equation of state and to the
dissipative, non-equilibrium processes. In the end one compare the predicted
distributions with those observed in experiments.

The hot and dense relativistic nuclear matter can appear in the form of 
hadronic matter or quark matter. While the hot matter might have existed in the
early universe the dense matter is supposed to be found in the interior of
neutron stars. The hot and dense nuclear matter is created in the relativistic
heavy ion collisions such as those at RHIC and those at LHC. 
When the space-time evolution of hot and dense matter is described by
dissipative, non-equilibrium fluid dynamics as is done in
\cite{Muronga04I,MR,Heinz,Romatschke,Muronga07I,Koide} knowledge of the equation of
state and transport coefficients is essential.

While the equation of state of the relativistic nuclear matter relates the state
variables the transport coefficients determine the strength of deviations from
equilibrium. The transport coefficients govern the dynamics of relaxation of the
matter towards equilibrium state.

Transport processes play an important role in the burning of neutron star into
a strange strange quark matter as described in \cite{Heiselberg}. When a
neutron at the phase boundary enters the quark matter, the quarks become
deconfined into a $u$ and two $d$ quarks. The $d$ quarks are transported into 
the burning region where they are converted to $s$
quarks. The $s$ quarks are then transported to the
phase boundary or the burning front. The $d$ and $s$ quarks diffuse through
each other and the $u$ quarks. This generate flows of different quark
flavors relative to each other. The burning drives the diffusion which which
is balanced by friction due to collisions. In the hadronic mixtures as
discussed in \cite{Prakash} transport properties plays an important role in
determining the relaxation times for different components. The viscosities and
thermal conductivities in fluid mixtures, e.g., pion-kaon $\pi K$ or quark-gluon
$q g$  mixture, should be larger than those in single component fluids, e.g,
$\pi\pi$, $qq$ or $gg$. In the mixture the diffusion coefficients depend not
only on the total density but also on the concentrations.

\section{The equation of state of hot and dense nuclear matter}
%%%%%%%%%%%%%%%%%%%%%%%%%%%%%%%%%%%%%%%%%%%%%%%%%%%%%%%%%%%%%%
\begin{figure}[ht]
\vspace{1.5cm}
\centerline{
     \includegraphics[width=11cm,height=10cm]{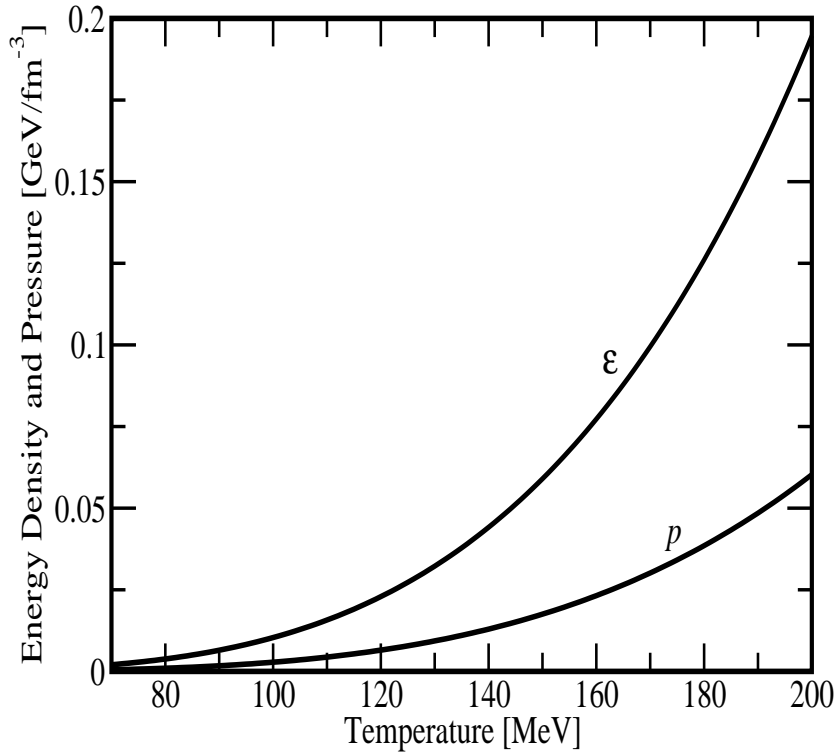}}
      \caption{The eaution of state: energy density and pressure as functions 
      of temperature for pure pionic matter.}
        \label{fig:eos}
\end{figure}
%%%%%%%%%%%%%%%%%%%%%%%%%%%%%%%%%%%%%%%%%%%%%%%%%%%%%%%%%%%%%%
The equation of state of relativistic nuclear matter is not that well known. The
equation of state for hadronic matter is often approximated by an equation of
state of free resonance gas. The hadronic equation of state is also affected
by the in-medium modifications. The equation of state of quark matter is
obtained from lattice QCD.

In this work the equation of state is taken to be that of resonance gas of
pions only. The energy density, number density and pressure are respectively
given by 
\begin{eqnarray}
\label{eq:eos}
  \label{eq:eoset}
  \varepsilon(T,\mu) & = &
  \sum_{k} g_{k}\int\frac{d^{3}p}{(2\pi)^{3}}
  \frac{ E_k}{e^{\frac{E_k-\mu}{T}} - 1},\\
  \label{eq:epstemp}
  n(T,\mu) & = &
  \sum_{k} g_{k}\int\frac{d^{3}p}{(2\pi)^{3}}
  \frac{1}{e^{\frac{E_k-\mu}{T}} - 1},\\
  \label{eq:ntemp}
  p(T,\mu) & = &
  \sum_{k} g_{k}\int\frac{d^{3}p}{(2\pi)^{3}}
  \frac{p^{2}}{3E_k}
  \frac{1}{e^{\frac{E_k-\mu}{T}} - 1}, \label{eq:presstemp} 
\end{eqnarray}
where $g_{k}$ is a degeneracy factor. In these calculations  
the pion chemical potential $\mu$ is fixed to zero.
In Fig.~\ref{fig:eos} we show the equation of state, i.e., the energy density
and pressure as functions of temperature. In the next section we will show that
the ratios of the standard transport coefficients to their corresponding
relaxation times is governed by the equation of state. This has a consequence
that the theory that gives the equation of state should also be able to yield
the transport coefficients if one wishes to calculate the two
self-consistently.

\section{Transport coefficients of hot and dense nuclear matter}
                                                                                                                                             
The transport coefficients are generally calculated using the relativistic
kinetic theory and thereby imply the knowledge of a collision term. The
relaxation times are considered to be given by more sophisticated theory or
evaluated roughly, so that we compare the standard transport coefficients (
thermal conductivity $\kappa$, bulk viscosity $\zeta$ and shear viscosity
$\eta$) divided
by their associated relaxation times (these are the relaxation times for the 
heat flux  $\tau_q$, for the bulk
viscous pressure  $\tau_\Pi$ and for the shear viscous pressure $\tau_\pi$). For
the thermal conductivity we will use $\lambda = \kappa T$ which is the quantity
that enters the dissipative fluid dynamic equations \cite{Muronga07I}. Thus we
will consider the ratios $\lambda/\tau_q$, $\zeta/\tau_\Pi$ and  $\eta/\tau_\pi$.
For the sake of brevity, only the calculations performed for pure pionic matter
are presented in the results. From Ref. \cite{Muronga07II} the ratios of
transport coefficients to their associated relaxation times are related to the
$\beta_A$ by
\begin{eqnarray}
{\zeta\over \tau_\Pi} &=& {1\over \beta_0}~,\\
{\lambda\over \tau_q}&=& {\kappa T\over \tau_q} = {1\over \beta_1}~,\\
{2\eta \over \tau_\pi} &=& {1\over \beta_2}~,
\end{eqnarray}
The second order coefficients, i.e., $\beta_0(\varepsilon, n)$,
$\beta_1(\varepsilon, n)$ and $\beta_2(\varepsilon, n)$ are given
by the equation of state. They are all positive and they make the second order
theories of relativistic fluid dynamics causal \cite{Muronga07I}.  They are
presented in detail in Ref. \cite{Muronga07II}. Thus the ratios of the
transport coefficients to relaxation times should be governed by the equation of
state.

\begin{figure}[hb]
\vspace{1.3cm}
\centerline{
     \includegraphics[width=11cm,height=11cm]{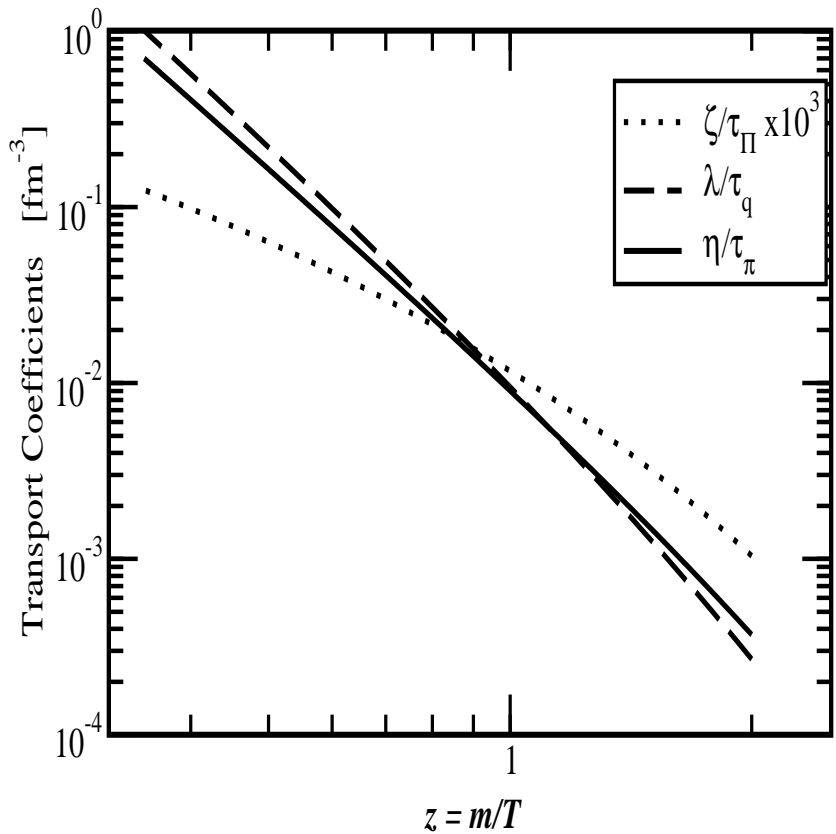}}
      \caption{Transport coefficients $\zeta/\tau_\Pi$, $\lambda/\tau_q$, 
      $\eta/\tau_\pi$, as functions of the
      "temperature parameter" $z=m/T$ for pure pionic matter.}
        \label{fig:transport}
\end{figure}

In Fig.~\ref{fig:transport} we show the transport coefficients  divided by
their respective relaxation times, i.e., $\lambda/\tau_q$, $\zeta/\tau_\Pi$ and
$\eta/\tau_\pi$.                                                                                                                                                                                                              
%%%%%%%%%%%%%%%%%%%%%%%%%%%%%%%%%%%%%%%%%%%%%%%%%%%%%%%%%%%%%%

The bulk viscosity of the pionic matter is very small. In
Fig.~\ref{fig:transport} it is  magnified one thousand times in order to compare
its temperature dependence with the other two coefficients. The thermal
conductivity times temperature, i.e., $\lambda$ is comparable to the shear
viscosity. Since the transport coefficients of the pionic matter increases with
temperature and the relaxation times decreases with temperature
\cite{Prakash, Muronga04II} one would expect the ratio of these transport
coefficients to their associated relaxation times to increase with rising
temperature. The $\beta_A$ are inversely proportional to the pressure.  The
ratios of the transport coefficients to their associated relaxation times are
inversely proportional to the $\beta_A$ and thus directly proportional to the
pressure. Hence like the pressure the ratios of the transport coefficients to
their associated relaxation times increases with temperature.

%%%%%%%%%%%%%%%%%%%%%%%%%%%%%%%%%%%%%%%%%%%%%%%%%%%%%%%%%%%%%%%%%%%%%%%%%%%%%%

\section{Conclusions and outlook}

A specific model has to be chosen to describe the space-time evolution of
relativistic nuclear matter. Using relativistic dissipative fluid dynamics
necessitates the knowledge of the underlying equation of state and the
transport coefficients or relaxation times for dissipative fluxes.  It has been
shown that the model/theory that gives the equation of state of the system
under consideration should be the same that gives the transport coefficients of
the system for consistency. In our calculations we considered a $\pi\pi$
system. It is clear from this simple study that the transport coefficients are
as important as the equation of state.

In this work we focused on the three standard transport coefficients, namely
the thermal conductivity, the bulk viscosity and the shear viscosity. This is
relevant in the present case study of a single component system: the $\pi\pi$
system. When one studies a multi-component mixtures the diffusion coefficients
become  important, e.g., flavor diffusion. In the case of flavor diffusion the
particle flavors are initially separated spatially. That is, the flavor chemical
potential depends on position. The flavor will be flowing with an associated
flow velocity. That is, we can define a drift velocity for each component of
the mixture independently of the four-flow velocity as a whole. For example in
a pion-kaon $\pi K$ mixture the kaons will flow with a certain velocity and
they will drift through the pions. In the case of a mixture of more than two
components the calculations become complex and one might resort to microscopic
transport models \cite{Muronga04II, Sasaki}.

%%%%%%%%%%%%%%%%%%%%%%%%%%%%%%%%%%%%%%%%%%%%%%%%%%%%%%%%%%%%%%%%

\end{document}